\documentclass[reprint,nofootinbib,showpacs]{revtex4}
\usepackage{graphicx}  
\usepackage{dcolumn}   
\usepackage{bm}        
\usepackage{amssymb}
\usepackage{hyperref}

\begin{document}

\title{\bf Signature of Lepton flavor universality violation in $B_s \to D_s\tau\nu$ semileptonic decays}
\author{Rupak~Dutta${}$}
\email{rupak@phy.nits.ac.in}
\author{Rajeev.~N${}$}
\email{16-3-24-102@student.nits.ac.in} 
\affiliation{
National Institute of Technology Silchar, Silchar 788010, India\\
}

\begin{abstract}
Deviation from the standard model prediction is observed in many semileptonic $B$ decays mediated via $b \to c$ charged current 
interactions. In particular, current experimental measurements of the ratio of branching ratio $R_D$ and $R_{D^{\ast}}$ in 
$B \rightarrow D^{(*)}l \nu$ decays disagree with standard model expectations at the level of about $4.1\sigma$. Moreover, recent 
measurement of the ratio of
branching ratio $R_{J/\Psi}$ by LHCb, where $R_{J/\Psi} = \mathcal B(B_c \to J/\Psi\,\tau\nu)/\mathcal B(B_c \to J/\Psi\,\mu\nu)$, is more
than $2\sigma$ away from the standard model prediction. In this context,
we consider an effective Lagrangian in the presence of vector and scalar new physics couplings to study the 
implications of $R_D$ and $R_{D^{\ast}}$ anomalies in $B_s \to D_s\,\tau\nu$ decays.  
We give prediction of several observables such as branching ratio, ratio of branching ratio, forward backward asymmetry parameter, $\tau$
polarization fraction, and the convexity parameter for the $B_s \to D_s\,\tau\nu$ decays within the standard model and within various
new physics scenarios.
\end{abstract}

\maketitle

\section{Introduction}
\label{int}
There are several reasons to believe that standard model (SM) of particle physics is not a complete theory, and thus there must be physics 
beyond the SM. It is therefore crucial to find the pattern of the New Physics (NP) that is responsible for various long standing anomalies.
The underlying framework of SM assumes that the charge and neutral leptons are universal in the weak interaction. However, various recent 
studies on semileptonic $B$ decays such as $B \rightarrow D^{(*)}l \nu$, with $l$ either $e$, $\mu$, or $\tau$, challenged the lepton flavor 
universality~\cite{Ciezarek:2017yzh}.
From past few years, many experiments such as $B-$factories have reported observables that are deviating from the SM prediction. 
In particular, the ratio of branching ratio $R_D$ and $R_{D^{\ast}}$ in $B \rightarrow D^{(*)}l \nu$ are measured to have large discrepancy
with respect to its SM counterpart.

A very precise SM prediction of the ratio of branching ratio $R_D$ in $B \rightarrow D\,l\,\nu$ using the form factors obtained 
in lattice quantum chromodynamics~(QCD) is reported
to be $0.300 \pm 0.008$~\cite{Lattice:2015rga,Na:2015kha,Aoki:2016frl,Bigi:2016mdz}. Similarly for $R_{D^{\ast}}$, it was reported to be
$0.252 \pm 0.003$~\cite{Fajfer:2012vx}. Comparing with the current world average of 
$R_D = 0.403 \pm 0.040 \pm 0.024$ and $R_{D^*} = 0.310 \pm 0.015 \pm 0.008$ from BABAR~\cite{Lees:2013uzd},
Belle~\cite{Huschle:2015rga,Sato:2016svk,Hirose:2016wfn}, and LHCb~\cite{Aaij:2015yra}, the combined deviation currently stands at about 
$4.1\sigma$. For definiteness, we report in Table-\ref{tabavg} the current status of
experimentally measured ratio of branching ratio $R_D$ and $R_{D^{\ast}}$~\cite{Amhis:2016xyh}.
\begin{table}[b]
\centering
\begin{tabular}{|c|c|c|}
\hline
Experiments & $ R_{D^{\ast}} $ & $R_D$ \\
\hline
\hline
BABAR &$0.332\pm 0.024\pm 0.018$ &$0.440 \pm 0.058 \pm 0.042 $ \\
BELLE &$0.293 \pm 0.038 \pm 0.015$ &$0.375 \pm 0.064 \pm 0.026$ \\
BELLE &$0.302 \pm 0.030 \pm 0.011 $ & \\
LHCb &$0.336 \pm 0.027 \pm 0.030 $ & \\
BELLE &$0.270 \pm 0.035^{+ 0.028}_{-0.025}$ & \\
LHCb & $0.285 \pm 0.019 \pm 0.029$ & \\
\hline
\hline
AVERAGE &$0.304 \pm 0.013 \pm 0.007$ &$0.407 \pm 0.039 \pm 0.024 $ \\
\hline
\end{tabular}
\caption{Current status of $R_D$ and $R_{D^{\ast}}$~\cite{Amhis:2016xyh}.}
\label{tabavg}
\end{table}
Various studies in explaining the observed anomalies in $B$ meson decays can be found in~\cite{Dutta:2013qaa,Celis:2012dk,Dutta:2016eml,
Alok:2016qyh,Ivanov:2016qtw,Tran:2017udy,Celis:2016azn,Dhargyal:2016nxq,Colangelo:2018cnj,Tanaka:1994ay,Nierste:2008qe,Miki:2002nz,
Fajfer:2012jt,Crivellin:2012ye,Datta:2012qk,Duraisamy:2013kcw,Duraisamy:2014sna,Biancofiore:2013ki,He:2012zp,Deshpand:2016cpw,Li:2016vvp,
Bardhan:2016uhr,Ivanov:2015tru,Nandi:2016wlp,Alonso:2016oyd,Altmannshofer:2017poe,Iguro:2017ysu,Bernlochner:2017jka,Bhattacharya:2016zcw}.
Very recently, LHCb has measured the ratio of branching ratio $R_{J/\psi}$ to be $0.71 \pm 0.17 \pm 0.18$~\cite{Aaij:2017tyk}. Comparing
with the SM prediction~\cite{Ivanov:2005fd,Wen-Fei:2013uea,Dutta:2017xmj}, we find the deviation to be at more than $2\sigma$. Although 
a precise calculation of $B_c \to J/\Psi$ form factors is not available at present, a preliminary results 
for the form factors are provided by HPQCD collaboration using Lattice QCD~\cite{Lytle:2016ixw}. 

Inspired by the anomalies present in $B \to (D,\,D^{\ast})\tau\nu$ decays, we study the corresponding $B_s \to D_s\tau\nu$ semileptonic
decays within SM and within various NP scenarios. A systematic study of $B_s \to D_s\tau\nu$ decays is important for several reasons: first,
in the limit of SU(3) flavor symmetry, $B_s \to D_s\tau\nu$ and $B \to D\tau\nu$ decay modes should show similar properties. Second, since
$B \to (D,\,D^{\ast})\tau\nu$ and $B_s \to D_s\tau\nu$ decays are mediated via $b \to c$ charged current interaction, hence anomalies present 
in $B \to (D,\,D^{\ast})\tau\nu$ should show up in $B_s \to D_s\tau\nu$ mode as well. Again, a combined analysis of $B$ and $B_s$ meson 
decays theoretically and experimentally
may help us to determine $|V_{cb}|$ with higher precision and this may also give some hints on our understanding of inclusive and exclusive
determination of $|V_{cb}|$. The semileptonic $B_s \to D_s\,l\,\nu$ decays has been
studied by various authors~\cite{Atoui:2013mqa,Atoui:2013zza,Bailey:2012rr,Zhao:2006at,Bhol:2014jta,Chen:2011ut,Li:2010bb,Fan:2013kqa,
Bazavov:2009bb,Na:2012kp,Monahan:2016qxu,Li:2009wq}. Within the SM, the branching ratio and ratio of branching ratio have been 
calculated using the form factors obtained from  
perturbative QCD (pQCD), Constituent Quark Meson (CQM) model, covariant light-front quark model, light cone sum rule, and 
more recently from Lattice QCD. Our main motivation here is to study the implications of $R_D$ and $R_{D^{\ast}}$ anomalies on 
$B_s \to D_s\tau\nu$ decays in a model independent
way. To this end, we use an effective theory formalism in the presence of NP to give prediction on various
observables such as the decay rate, ratio of branching ratio, lepton side forward backward asymmetry, longitudinal polarization
fraction of the lepton,  and the convexity parameter for the $B_s \to D_s\tau\nu$ decays. We follow Ref.~\cite{Monahan:2017uby} for various $B_s \to D_s$
transition form factors. For our NP analysis, we impose $1\sigma$ constraint coming from the measured ratio of branching ratio $R_D$ and 
$R_{D^{\ast}}$ to obtain the allowed NP parameter space. This is to ensure that the resulting NP parameter space can simultaneously explain
the anomalies present in $R_D$ and $R_{D^{\ast}}$. We also use the constraint coming from the $B_c$ meson life time in our analysis as it is 
shown in Ref.~\cite{Alonso:2016oyd} that life 
time of $B_c$ meson put severe constraint on various NP couplings. Based on
various SM calculation~\cite{Bigi:1995fs,Beneke:1996xe,Chang:2000ac} of the $B_c$ life time, it is found that $\mathcal B(B_c \to \tau\nu)$ can not 
be more than $5\%$. However, this can be relaxed up to 
$30\%$ depending on the input parameters that are used in the SM calculation. Very recently, in Ref.~\cite{Akeroyd:2017mhr} a more stringent bound of 
$\mathcal B(B_c \to \tau\nu) \le 10\%$ was obtained using the LEP data at the $Z$ peak. 

The present discussion is organized as follows. In section.~\ref{pheno}, we start with a brief discussion of the most general effective
Lagrangian in the presence of NP for the $b \to c\,l\,\nu$ quark level transition decays. The three body differential decay width formula 
obtained using helicity formalism is also reported in the section.~\ref{pheno}. Explicit formulas of various observables such as ratio 
of branching ratio, lepton side forward backward asymmetry, $\tau$ polarization fraction, and the convexity parameter for the 
$B_s \to D_s\tau\nu$ decays in the presence of NP are reported.
In section.~\ref{rd}, we first report all the input parameters that are relevant for our numerical computation. Standard model results and 
the effects of NP couplings on various observables under various NP scenarios are reported in section.~\ref{rd}. Finally, we conclude and 
summarize our results in section.~\ref{con}.

\section{Phenomenology}
\label{pheno}
The natural way to introduce NP effects in a model independent approach is to construct a effective Lagrangian
for the weak decays that includes both SM and the beyond the SM physics. We follow Refs.~\cite{Cirigliano:2009wk,Bhattacharya:2011qm} and
write the effective weak Lagrangian for the $b \to c\,\tau\nu$ quark level transition decays in the presence of vector and scalar type NP
interactions as 
\begin{eqnarray}
\label{effl}
\mathcal L_{\rm eff} &=&
-\frac{4\,G_F}{\sqrt{2}}\,V_{c b}\,\Bigg\{(1 + V_L)\,\bar{l}_L\,\gamma_{\mu}\,\nu_L\,\bar{c}_L\,\gamma^{\mu}\,b_L +
V_R\,\bar{l}_L\,\gamma_{\mu}\,\nu_L\,\bar{c}_R\,\gamma^{\mu}\,b_R 
+
\widetilde{V}_L\,\bar{l}_R\,\gamma_{\mu}\,\nu_R\,\bar{c}_L\,\gamma^{\mu}\,b_L \nonumber \\
&&+
\widetilde{V}_R\,\bar{l}_R\,\gamma_{\mu}\,\nu_R\,\bar{c}_R\,\gamma^{\mu}\,b_R +
S_L\,\bar{l}_R\,\nu_L\,\bar{c}_R\,b_L +
S_R\,\bar{l}_R\,\nu_L\,\bar{c}_L\,b_R +
\widetilde{S}_L\,\bar{l}_L\,\nu_R\,\bar{c}_R\,b_L +
\widetilde{S}_R\,\bar{l}_L\,\nu_R\,\bar{c}_L\,b_R\Bigg\} + {\rm h.c.}\,,
\end{eqnarray}
where, $G_F$ is the Fermi coupling constant and $|V_{cb}|$ is the Cabibbo-Kobayashi-Mashkawa~(CKM) matrix element. The effective
Lagrangian of Eq.~\ref{effl} is considered at renormalization scale $\mu = m_b$. The NP Wilson coefficients~(WCs) denoted by 
$V_L$, $V_R$, $S_L$, and $S_R$ 
involve left-handed neutrinos, whereas, the WCs denoted by $\tilde{V_L}$, $\tilde{V_R}$, $\tilde{S_L}$, and $\tilde{S_R}$ involve 
right-handed neutrinos, respectively. Assuming all NP WCs to be real in the present analysis we rewrite the above equation 
as~\cite{Dutta:2013qaa},
\begin{eqnarray}
\label{leff}
\mathcal L_{\rm eff} &=&
-\frac{G_F}{\sqrt{2}}\,V_{c b}\,\Bigg\{G_V\,\bar{l}\,\gamma_{\mu}\,(1 - \gamma_5)\,\nu_l\,\bar{c}\,\gamma^{\mu}\,b -
G_A\,\bar{l}\,\gamma_{\mu}\,(1 - \gamma_5)\,\nu_l\,\bar{c}\,\gamma^{\mu}\,\gamma_5\,b +
G_S\,\bar{l}\,(1 - \gamma_5)\,\nu_l\,\bar{c}\,b \nonumber \\
&& - G_P\,\bar{l}\,(1 - \gamma_5)\,\nu_l\,\bar{c}\,\gamma_5\,b + 
\widetilde{G}_V\,\bar{l}\,\gamma_{\mu}\,(1 + \gamma_5)\,\nu_l\,\bar{c}\,\gamma^{\mu}\,b -
\widetilde{G}_A\,\bar{l}\,\gamma_{\mu}\,(1 + \gamma_5)\,\nu_l\,\bar{c}\,\gamma^{\mu}\,\gamma_5\,b \nonumber \\
&&+
\widetilde{G}_S\,\bar{l}\,(1 + \gamma_5)\,\nu_l\,\bar{c}\,b - \widetilde{G}_P\,\bar{l}\,(1 + \gamma_5)\,\nu_l\,\bar{c}\,
\gamma_5\,b\Bigg\} + {\rm h.c.}\,,
\end{eqnarray}
where, 
\begin{eqnarray} 
&&G_V = 1 + V_L + V_R\,,\qquad\qquad
G_A = 1 + V_L - V_R\,, \qquad\qquad
G_S = S_L + S_R\,,\qquad\qquad
G_P = S_L - S_R\,, \nonumber \\
&&\widetilde{G}_V = \widetilde{V}_L + \widetilde{V}_R\,,\qquad\qquad
\widetilde{G}_A = \widetilde{V}_L - \widetilde{V}_R\,, \qquad\qquad
\widetilde{G}_S = \widetilde{S}_L + \widetilde{S}_R\,,\qquad\qquad
\widetilde{G}_P = \widetilde{S}_L - \widetilde{S}_R\,.
\end{eqnarray}

The $B_s \to D_s\,l\,\nu$ decay amplitude depends on non perturbative hadronic matrix element which can be parametrized in terms of
$B_s \to D_s$ transition form factors as follows.
\begin{eqnarray}
&&\langle D_s (P_{D_s})|\bar{c}\,\gamma^{\mu}\,b|B_s (P_{B_s}) \rangle = f_+ (q^2) \left[ P^{\mu}_{B_s} + P^{\mu}_{D_s} - 
\frac{M_{B_{s}}^{2}-M_{D_s}^{2}}{q^{2}}q^{\mu} \right] + f_0(q^2) \frac{M_{B_{s}}^{2}-M_{D_s}^{2}}{q^{2}} q^{\mu}\,, \nonumber \\
&&\langle D_s (P_{D_s})|\bar{c}\,b|B_s (P_{B_s}) \rangle = \frac{m_{B_s}^2 - m_{D_s}^2}{m_b(\mu) - m_c(\mu)}\,f_0(q^2)\,,
\end{eqnarray}
where, $q^{\mu}=P_{B_s}^{\mu}-P_{D_s}^{\mu}$ refers to the momentum transfer. It should be mentioned that we use the equation of motion to
find the scalar matrix element. We follow Ref.~\cite{Monahan:2017uby} for the relevant form factors $f_0(q^2)$ and $f_+(q^2)$. The 
expressions pertinent for our discussion are~\cite{Monahan:2017uby}
\begin{eqnarray}
&&P_{0}(q^{2}) f_{0}(q^{2})= \sum_{j=0}^{3} a_{j}^{(0)}\,z^j\,, \qquad\qquad
P_{+}(q^{2})f_{+}(q^{2})= \sum_{j=0}^{3} a_{j}^{(+)}\,\Big[z^{j}-(-1)^{j-J}\frac{j}{J}z^{J}\Big]
\end{eqnarray}
where
\begin{eqnarray}
&&z(q^2)=\frac{\sqrt{t_{+}-q^2}-\sqrt{t_{+}-t_0}}{\sqrt{t_{+}-q^2}+\sqrt{t_{+}-t_0}}\,\qquad 
t_{+}=(M_{B_s}+M_{D_s})^2\,\qquad
t_{0}=(M_{B_s}-M_{D_s})^2\,\qquad  
P_{0,+}(q^2)=1-\frac{q^2}{M_{0,+}^2}\,.
\end{eqnarray}
Here, $P_{0,+}$ are Blaschke factors and $M_{0}=6.42(10)\,{\rm GeV}$ and $M_{+}=6.330(9)\,{\rm GeV}$ are the resonance masses. We refer to 
Ref.~\cite{Monahan:2017uby} for all the omitted details.

The differential decay distribution for the $B_s \to D_s\,l\,\nu$ decays can be expressed as
\begin{eqnarray}
\frac{d\Gamma}{dq^2\,d\cos\theta} = \frac{G_F^2\,|V_{cb}|^2\,|\vec{P}_{D_s}|}{(2\pi)^3\,64\,m_{B_s}^2}\,\Big(1-\frac{m_l^2}{q^2}\Big)\,
L_{\mu\nu}\,H^{\mu\nu}\,,
\end{eqnarray}
where $|\vec{P}_{D_s}| = \sqrt{\lambda(m_{B_s}^2,\,m_{D_s}^2,\,q^2)}/2\,m_{B_s}$ is the three momentum vector of the outgoing meson 
and $\lambda(a,\,b,\,c) = a^2 + b^2 + c^2 - 2\,(a\,b + b\,c + c\,a)$. Note that $\theta$ denotes the angle between the $D_s$ meson and 
the lepton three momentum vector in the $(l\,\nu)$ rest frame. The covariant contraction $L_{\mu\nu}H^{\mu\nu}$ can be calculated using the
helicity techniques of Refs.~\cite{Korner:1989qb,Kadeer:2005aq}. For completeness, we present here the final expression for the
differential decay distribution of three body $B_s \to D_s\,l\,\nu$ decays~\cite{Dutta:2013qaa}. 
\begin{eqnarray}
\label{dslnutheta}
\frac{d\Gamma}{dq^2\,d\cos\theta} = 2\,N\,|\vec{P}_{D_s}|\,\Bigg\{H_0^2\,\sin^2\theta\,(G_V^2 + \widetilde{G}_V^2) + \frac{m_l^2}{q^2}\,
\Big[(H_0\,G_V\,\cos\theta - H_{tS})^2 + (H_0\,\widetilde{G}_V\,\cos\theta - \widetilde{H}_{tS})^2\Big]\Bigg\}\,,
\end{eqnarray}
where
\begin{eqnarray}
&&N = \frac{G_F^2\,|V_{c\, b}|^2\,q^2}{256\,\pi^3\,m_{B_s}^2}\,\Big(1 - \frac{m_l^2}{q^2}\Big)^2\,, \qquad\qquad
H_0 = \frac{2\,m_{B_s}\,|\vec{P}_{D_s}|}{\sqrt{q^2}}\,f_{+}(q^2) \qquad\qquad
H_t = \frac{m_{B_s}^2 - m_{D_s}^2}{\sqrt{q^2}}\,f_0(q^2)\,, \nonumber \\
&&H_S=\frac{m_{B_s}^2 - m_{D_s}^2}{m_b(\mu) - m_{c}(\mu)}\,f_0(q^2)\,,\qquad\qquad
H_{tS} = H_t\,G_V + \frac{\sqrt{q^2}}{m_l}\,H_S\,G_S\,,\qquad\qquad
\widetilde{H}_{tS} = H_t\,\widetilde{G}_V + \frac{\sqrt{q^2}}{m_l}\,H_S\,\widetilde{G}_S\,.
\end{eqnarray}
By performing the $\cos\theta$ integration in Eq.~\ref{dslnutheta}, we get
\begin{eqnarray}
\label{Dslnu}
\frac{d\Gamma}{dq^2} &=&
\frac{8\,N\,|\vec{P}_{D_s}|\,}{3}\Bigg\{\,H_0^2\,\Big(G_V^2 + \widetilde{G}_V^2\Big)\,\Big(1 + \frac{m_l^2}{2\,q^2}\Big) 
 + \frac{3\,m_l^2}{2\,q^2}\,\Big( H_{tS}^2 + \widetilde{H}_{tS}^2\Big) \Bigg\}\,.
\end{eqnarray}
Setting $G_V=G_A=1$ and all other NP couplings to zero, we obtain
\begin{eqnarray}
\Big(\frac{d\Gamma}{dq^2}\Big)_{\rm SM} = \frac{8\,N\,|\vec{P}_{D_s}|\,}{3}\Bigg\{\,H_0^2\,\Big(1 + \frac{m_l^2}{2\,q^2}\Big) + 
\frac{3\,m_l^2}{2\,q^2}\,H_t^2\Bigg\}\,.
\end{eqnarray}
We define several $q^2$ dependent observables such as differential branching ratio ${\rm DBR}(q^2)$, ratio of branching ratio $R(q^2)$, 
lepton side forward backward asymmetry $A_{FB}^l(q^2)$, polarization fraction of the charged lepton $P_{l}(q^2)$, and convexity 
parameter $C_F^l(q^2)$ for the $B_s \to D_s\,l\,\nu$ decays. Those are 
\begin{eqnarray}
&&{\rm DBR}(q^2)=\frac{d\Gamma/dq^2}{\Gamma_{\rm Tot}}\,, \qquad\qquad
R(q^2)=\frac{\mathcal B(B_s \rightarrow D_s \tau \nu)}{\mathcal B(B_s \rightarrow D_s\,l\,\nu)}\,, \qquad\qquad
A_{FB}(q^2) = \frac{\Big(\int_{-1}^{0}-\int_{0}^{1}\Big)d\cos\theta\frac{d\Gamma}{dq^2\,d\cos\theta}}{\frac{d\Gamma}{dq^2}} \,, \nonumber \\
&&P_{l}(q^2)=\frac{d\Gamma(+)/dq^2 - d\Gamma(-)/dq^2}{d\Gamma(+)/dq^2 + d\Gamma(-)/dq^2}\,, \qquad\qquad
C_{F}^{l}(q^2)= \frac{1}{\left(d\Gamma/dq^2\right)} \left(\frac{d}{d\cos\theta}\right)^2 \left[\frac{d\Gamma}{dq^2\,d\cos\theta}\right]\,,
\end{eqnarray}
where $d\Gamma(+)/dq^2$ and $d\Gamma(-)/dq^2$ represent differential decay width of positive and negative helicity leptons, respectively.
In the presence of various NP, the explicit expressions for $A_{FB}^l$, $d\Gamma(+)/dq^2$, $d\Gamma(-)/dq^2$, and $C_F^l$ are
\begin{eqnarray}
\label{obs}
&&A_{FB}^l (q^2)= \frac{3m_{l}^{2}}{2q^2}\frac{H_0\,G_V\,H_{tS} + H_0\,\widetilde{G}_V\,\widetilde{H}_{tS}}
{H_0^2\,(G_V^2+\widetilde{G}_V^2)\Big(1+\frac{m_l^2}{q^2}\Big)+\frac{3\,m_l^2}{2q^2}\Big(H_{tS}^2 + \widetilde{H}_{tS}^2\Big)}\,,\nonumber \\
&&\frac{d\Gamma(+)}{dq^2}= \frac{8\,N|\vec{P_{D_s}}|}{3} \Big[H_{0}^{2}\,\widetilde{G}_V^2 + \frac{m_l^2}{2q^2}\,\Big(H_0^2\,G_V^2+3\,
H_{tS}^2\Big)\Big]\,\nonumber \\
&&\frac{d\Gamma(-)}{dq^2}= \frac{8\,N|\vec{P_{D_s}}|}{3}\Big[H_0^2\,G_V^2 + \frac{m_l^2}{2q^2}\Big(H_0^2\,\tilde{G}_V^2+3\,
\widetilde{H}_{tS}^2\Big)\Big]\,,\nonumber \\
&&C_F^{l}= \frac{3}{2}\frac{H_0^2\Big(G_V^2 + \widetilde{G}_V^2)\Big(\frac{m_l^2}{q^2}-1\Big)}
{\Big[H_0^2\,(G_V^2+\widetilde{G}_V^2)(1+\frac{m_l^2}{2\,q^2})+\frac{3\,m_l^2}{2\,q^2}(H_{tS}^2+\widetilde{H}_{tS}^{2})\Big]} 
\end{eqnarray}
The SM expressions are obtained by setting all the NP couplings to zero. 
\begin{eqnarray}
\label{smobs}
&&\Big[A_{FB}^l(q^2)\Big]_{SM}= \frac{3\,m_l^2}{2\,q^2}\Bigg\{\frac{H_0\,H_t}{H_0^2\,\Big(1+\frac{m_l^2}{q^2}\Big)+\frac{3\,m_l^2}{2\,q^2}\,
H_t^2}\Bigg\}\,, \qquad
\Big[P^l\Big]_{SM}= \frac{\frac{m_l^2}{2\,q^2}\,(H_0^2+3\,H_t^2) - H_0^2}{\frac{m_l^2}{2\,q^2}\,(H_0^2+3\,H_t^2) + H_0^2}\,, \nonumber \\
&&\Big[C_F^l\Big]_{SM}= \frac{3}{2}\frac{H_0^2\Big(\frac{m_l^2}{q^2}-1\Big)}
{\Big[H_0^2\,\Big(1+\frac{m_l^2}{2\,q^2}\Big)+\frac{3\,m_l^2}{2\,q^2}\,H_t^2\Big]}
\end{eqnarray}
The average values of the forward-backward asymmetry of the charged lepton $<A_{FB}^l>$, the longitudinal polarization fraction of the 
lepton $<P^l>$, and the convexity parameter $<C_F^l>$ are obtained by separately integrating the numerators and denominators over $q^2$.

\section{Numerical results and discussions}
\label{rd}
\subsection{Input parameters}
Before proceeding for the analysis, we report in Table-\ref{tabinp} all the input parameters that are relevant for our numerical computation.
For the mass and lifetime parameter, we use the latest values reported in Ref.~\cite{Patrignani:2016xqp}. Similarly for the CKM matrix element $|V_{cb}|$ and
the Fermi coupling constant $G_F$, we use Ref.~\cite{Patrignani:2016xqp}. 
The lepton masses~($m_e$, $m_{\tau}$) and meson masses~($B_s$, $D_s$) are in ${\rm GeV}$ units, whereas
Fermi coupling constant $G_F$ is in ${\rm GeV}^{-2}$ units. The lifetime of $B_s$ meson~($\tau_{B_s}$) is in seconds. The quark masses 
$m_b (m_b)$ and $m_c (m_b)$ evaluated at renormalization scale $\mu=m_b$ are in ${\rm GeV}$ units.
For the relevant $B_s \to D_s$ form factor parameters, we follow the most recent Lattice QCD calculation of Ref.~\cite{Monahan:2017uby}. The
uncertainties associated with $|V_{cb}|$, and the form factors parameters are written within parenthesis. We do not report the uncertainties
associated with other input parameters as they do not play an important role in our analysis.
\begin{table}[htbp]
 \centering
\begin{tabular}{ ll|ll }
 \hline
 \hline
  Inputs from PDG ~\cite{Patrignani:2016xqp} &  & Form factor inputs ~\cite{Monahan:2017uby} &  \\
 \hline
$m_{B_s}$ & 5.36689 &  $a_{0}^{(0)}$ & 0.658(31)  \\
$m_{D_s}$ & 1.96827 & $a_{1}^{(0)}$ & -0.10(30)   \\
$m_b (\mu)$ & 4.18 &  $a_{2}^{(0)}$ & 1.3(2.8)   \\
$m_c (\mu)$ & 0.91 & $a_{0}^{(+)}$ & 0.858(32)   \\
$m_e$ & 0.05109989461$\times 10^{-2}$ &  $a_{1}^{(+)}$ & -3.38(41)   \\
$m_{\tau}$ & 1.77682 &  $a_{2}^{(+)}$ & 0.6(4.7) \\
$|V_{cb}|$ & 0.0409(11) &            &          \\
$G_F$ & $1.1663787 \times 10^{-5}$ &  &  \\
$\tau_{B_s}$ & $1.505\times 10^{-12}$  & & \\
 \hline
 \hline
\end{tabular}
\caption{Theory inputs}
 \label{tabinp}
\end{table}

We wish to determine the consequences of various NP couplings on various observables for the $B_s \to D_s\tau\nu$ decays in a model
independent way. It is, therefore, crucial to determine the size of the SM uncertainties in each observable that may come from various
input parameters. Uncertainties in the theoretical prediction of the observables mainly come from two sources. First, it may come from
not very well known CKM matrix element $|V_{cb}|$ and second, it may come from the non perturbative hadronic inputs such as decay constant
and meson to meson form factors. To gauge the effect of above mentioned uncertainties on various observables, we use a random number
generator and vary these input parameters within $1\sigma$ of their central values. 

\subsection{Standard model prediction}
First, we wish to give prediction of various observables for both the $e$ and $\tau$ mode within the SM. 
We report in Table.-\ref{tabsm} the SM central values and the $1\sigma$ ranges of each observable for the $e$ and the $\tau$ modes.
Here, the central values are obtained by considering only the central values of theory input parameters whereas, the $1\sigma$ ranges of
each observable is obtained by performing a random scan of the hadronic parameters and the CKM matrix element within $1\sigma$ of their 
central values. The value of ratio of branching ratio $R_{D_s}$ in Table.-\ref{tabsm} is quite similar to the value reported in 
Ref.~\cite{Monahan:2017uby}. The slight difference may come from different choices of input parameters.
\begin{table}[htbp]
\centering
\begin{tabular}{|c|c|c||c|c|c||c|c|c|}
\hline
Observables & Central value &$1\sigma$ range &Observables &Central value &$1\sigma$ range & Observables &Central value &$1\sigma$ range\\
\hline
\hline
$\mathcal B(B_s \to D_s\,e\nu)\%$ &$2.238$ &$[1.839, 2.693]$ & $P^e$ &$-1.00$ &$-1.00$&$P^{\tau}$ &$0.320$ 
&$[0.234, 0.403]$\\
\hline
$\mathcal B(B_s \to D_s\tau\nu)\%$ &$0.670$ &$[0.573, 0.777]$ &$A^e_{FB}$ &$0.00$ &$0.00$&$A^{\tau}_{FB}$ &$0.360$ &$[0.352, 0.364]$\\
\hline
$R_{D_s}$ &$0.299$ &$[0.260, 0.351]$ &$C^e_F$ &$-1.5$ &$-1.50$&$C^{\tau}_F$ &$-0.271$ &$[-0.239, -0.305]$\\
\hline
\hline
\end{tabular}
\caption{SM prediction of various observables for the $e$ and the $\tau$ modes}
\label{tabsm}
\end{table}

\begin{figure}[htbp]
\centering
\includegraphics[width=6.0cm,height=4.3cm]{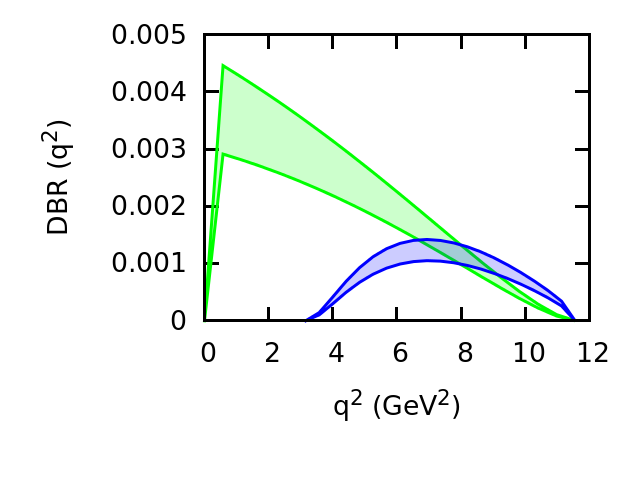}
\includegraphics[width=6.0cm,height=4.3cm]{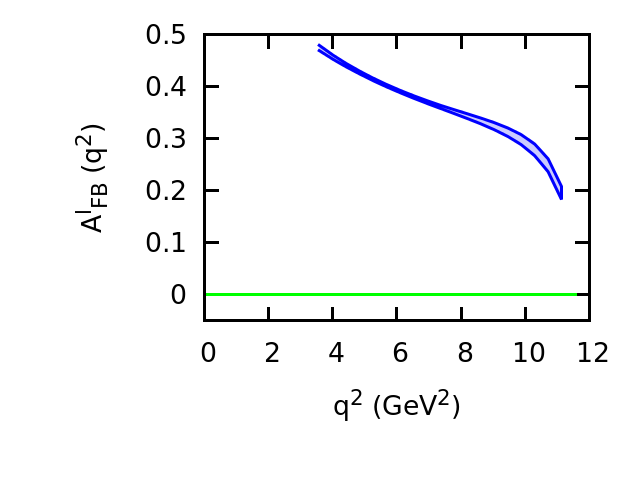}
\includegraphics[width=6.0cm,height=4.3cm]{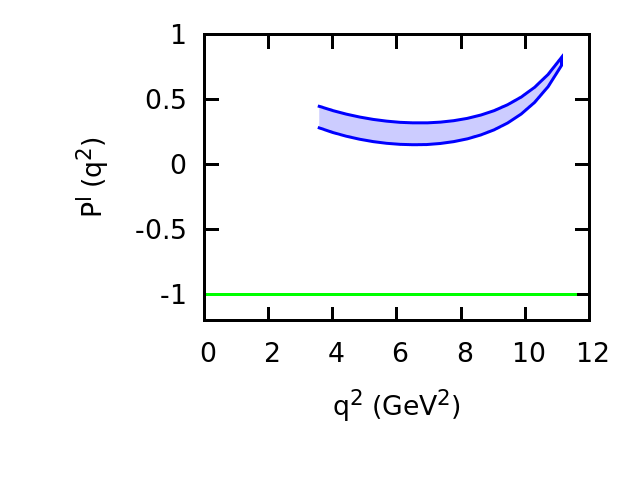}
\includegraphics[width=6.0cm,height=4.3cm]{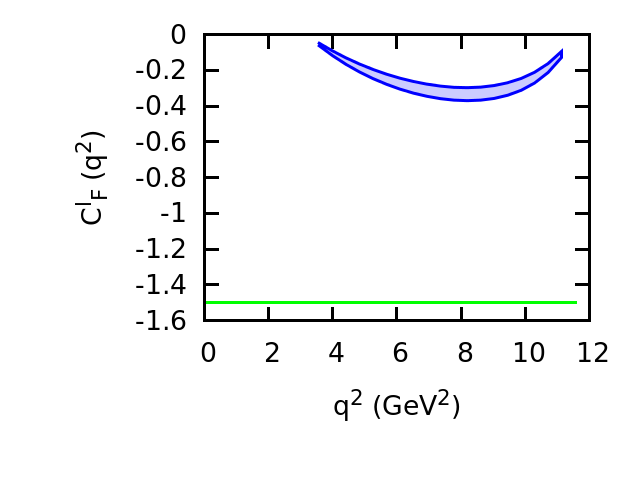}
\caption{$q^2$ dependence of various observables in the SM for the $e$~(green) and the $\tau$~(blue) modes.}
\label{obs_sm}
\end{figure}

We notice that the SM prediction for the $e$ mode is quite different from the $\tau$ mode. There is even a sign change in the polarization
fraction $P_l$ while going from the $e$ to the $\tau$ mode. Again, the forward backward asymmetry parameter for the $e$ mode is zero, whereas,
it is non zero positive for the $\tau$ mode. Similarly, the convexity parameter $C_F^l$ for the $e$ mode is much larger in magnitude than for
the $\tau$ mode. It is worth mentioning that the mass of the charged lepton plays an important role. In Fig.~\ref{obs_sm}, we show the $q^2$
dependence of each observable for the $e$ and the $\tau$ modes, respectively. We notice that the $A_{FB}^l(q^2)$, $P^l(q^2)$ and the 
$C_F^l(q^2)$ observables remain constant in the entire $q^2$ region for the $e$ mode. This could be very well understood from Eq.~\ref{smobs}.
In the massless limit, i.e, in the $m_l \to 0$ limit, the $q^2$ dependence gets cancelled in the ratio for the $A_{FB}^l(q^2)$, $P^l(q^2)$ 
and the $C_F^l(q^2)$ observables.

Now we proceed to discuss various NP effects in $B_s \to D_s\,\tau\,\nu$ decays.

\subsection{New physics in $B_s \to D_s\,\tau\,\nu$ decays}
Study of $B_s \to D_s\,\tau\,\nu$ decays both theoretically and experimentally is well motivated because of the long standing anomalies
present in $R_D$ and $R_{D^{\ast}}$. We wish to study the implication of these existing anomalies on the $B_s \to D_s\,\tau\,\nu$ decays
in a model independent way. We consider four different NP scenarios based on NP contributions from two different operators. In order to 
determine the allowed NP parameter space, we impose $1\sigma$ constraint coming from the measured ratio of branching ratios $R_D$ and 
$R_{D^{\ast}}$. We use the average values of $R_D$ and $R_{D^{\ast}}$ reported in Table.~\ref{tabavg} in our analysis. For the uncertainties
we added the statistical and systematic uncertainties in quadrature. Again, we assume that only the third generation leptons get contribution 
from NP. 
\subsubsection{Scenario I: only $V_L$ and $V_R$ NP couplings}
\label{vl_vr}
\begin{figure}[htbp]
\centering
\includegraphics[width=6.0cm,height=4.3cm]{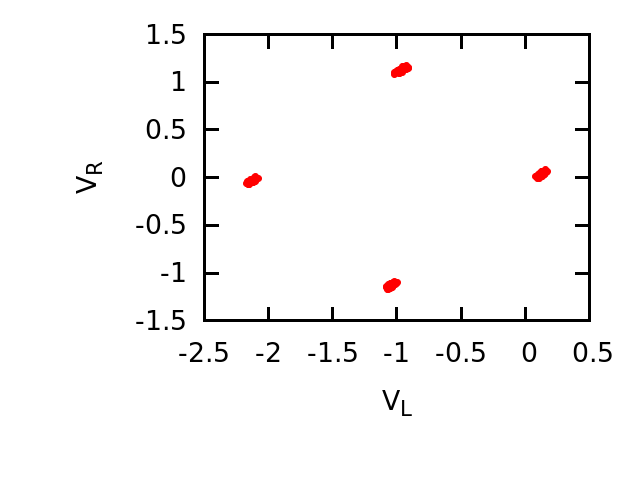}
\includegraphics[width=6.0cm,height=4.3cm]{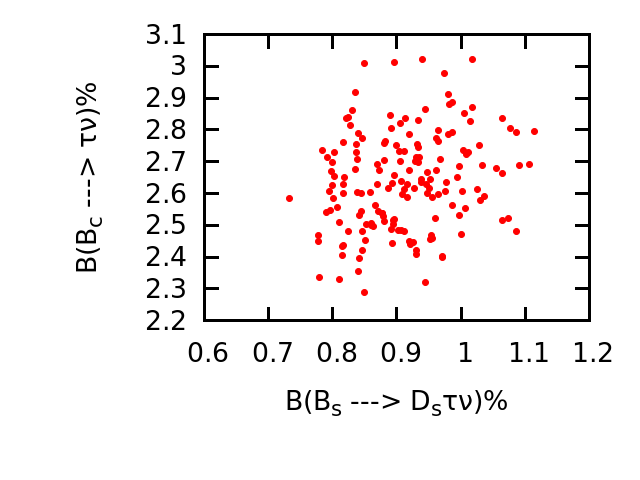}
\caption{Allowed ranges of $V_L$ and $V_R$ NP couplings are shown in the left panel once $1\sigma$ constraint coming from the measured
values of the ratio of branching ratios $R_D$ and
$R_{D^{\ast}}$ is imposed. We show in the right panel the allowed ranges in $\mathcal B(B_c \to \tau\nu)$ and $\mathcal B(B_s \to D_s\tau\nu)$
in the presence of these NP couplings.}
\label{vlvr}
\end{figure}

In this scenario, we vary $V_L$ and $V_R$ and set all other NP couplings to zero. This is to ensure that NP contribution to the 
$B_s \to D_s\tau\nu$ decay mode is coming only from vector type NP couplings that involves left handed neutrinos. 
In the presence of such NP, the $d\Gamma/dq^2$, $R(q^2)$, $A_{FB}^{\tau}(q^2)$, $P^{\tau}(q^2)$, and $C^{\tau}_F(q^2)$ can be 
expressed as
\begin{eqnarray}
\label{vlvr_eq}
&&\left[ \frac{d\Gamma}{dq^{2}} \right]_{V_{L,R}}= \left[ \frac{d \Gamma}{dq^{2}} \right]_{SM} G_V^2\,, \qquad\qquad
\left[R_{D_s} \right]_{V_{L.R}}=\left[R_{D_s} \right]_{SM} G_V^2\,, \nonumber \\
&& \left[A_{FB}^l(q^2)\right]_{V_{L,R}}= \left[A_{FB}^l(q^2)\right]_{SM}\,, \qquad\qquad
 \left[P^l\right]_{V_{L,R}}= \left[P^l\right]_{SM}\,, \qquad\qquad
 \left[C_F^l\right]_{V_{L,R}}= \left[C_F^l\right]_{SM}
\end{eqnarray}
It is evident from Eq.~\ref{vlvr_eq} that $d\Gamma/dq^2$ and $R(q^2)$ depend on $V_L$ and $V_R$ NP couplings and are proportional to $G_V^2$, 
whereas, $P^{\tau}(q^2)$, $A^{\tau}_{FB}(q^2)$, and $C^{\tau}_F(q^2)$ do not depend on these NP couplings since the contribution coming from
$V_L$ and $V_R$ NP couplings gets canceled in the ratio. The allowed ranges of
$V_L$ and $V_R$ after imposing $1\sigma$ constraint coming from $R_D$ and $R_{D^{\ast}}$ are shown in the left panel of Fig.~\ref{vlvr}.
In the right panel we show the corresponding ranges in $\mathcal B(B_c \to \tau\nu)$ and $\mathcal B(B_s \to D_s\tau\nu)$. From the right
panel of Fig.~\ref{vlvr}, we notice that the $\mathcal B(B_c \to \tau\nu)$ obtained in this scenario lies in the $2\% - 3\%$ range. This is 
consistent with the SM calculation. 
We report in Table-\ref{tabvlvr} the allowed ranges of each observable for the $B_s \rightarrow D_s \tau \nu$ decays with $(V_L$, $V_R)$ NP 
couplings of Fig.~\ref{vlvr}. We see significant deviation in $\mathcal B(B_s \to D_s\tau\nu)$ and $R_{D_s}$ from the SM prediction. As
expected, the ranges of $P^{\tau}_{D_s}$, $A^{\tau}_{FB}$, and $C^{\tau}_F$ do not vary at all with such NP couplings. 
\begin{table}[htbp]
\centering
\begin{tabular}{|c|c|c|c|c|}
\hline
$\mathcal B(B_s \to D_s\tau\nu)\%$ & $R_{D_s}$ & $P^{\tau}$ & $A^{\tau}_{FB}$ & $C^{\tau}_F$ \\
\hline
\hline
$[0.733, 1.115]$ & $[0.329, 0.496]$ & $[0.234, 0.403]$ & $[0.352, 0.364]$ & $[-0.239, -0.305]$ \\
\hline
\end{tabular}
\caption{Allowed ranges of various observables with $V_L$ and $V_R$ NP couplings of Fig.~\ref{vlvr}.}
\label{tabvlvr}
\end{table}
\begin{figure}[htbp]
\centering
\includegraphics[width=5.9cm,height=4.3cm]{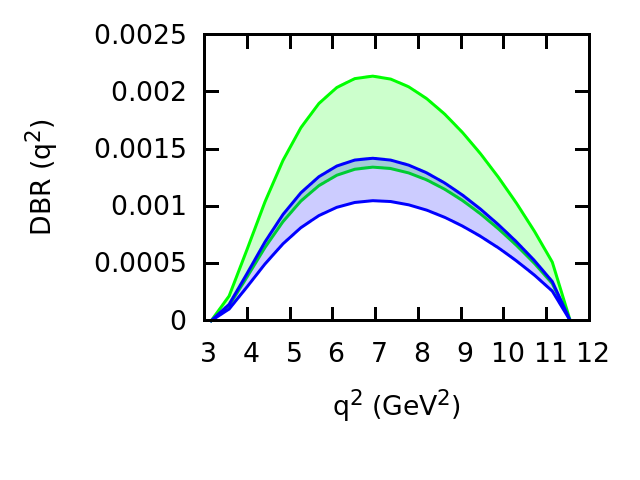}
\includegraphics[width=5.9cm,height=4.3cm]{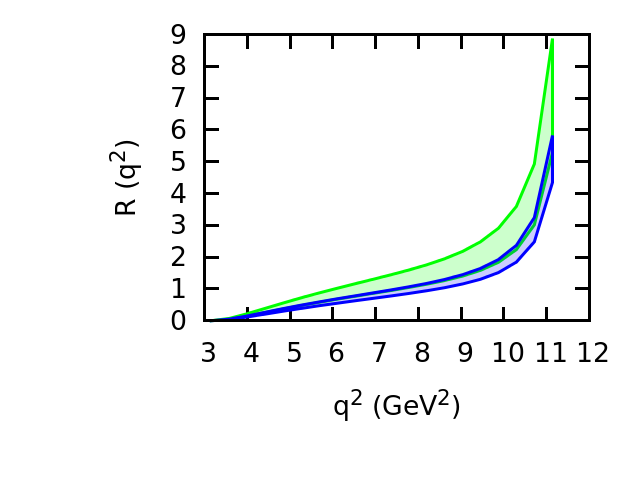}
\includegraphics[width=5.9cm,height=4.3cm]{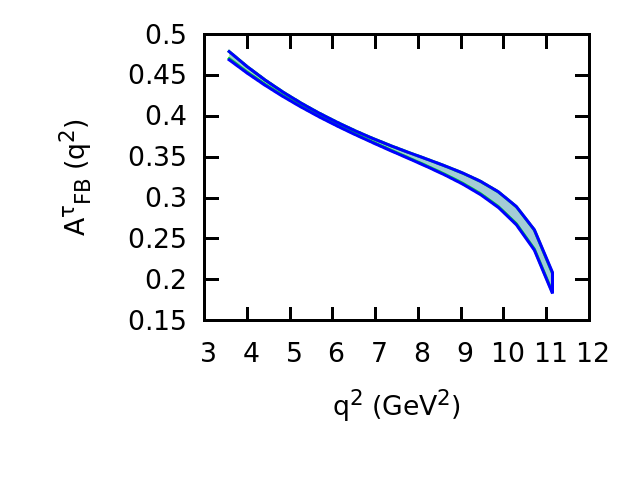}
\includegraphics[width=5.9cm,height=4.3cm]{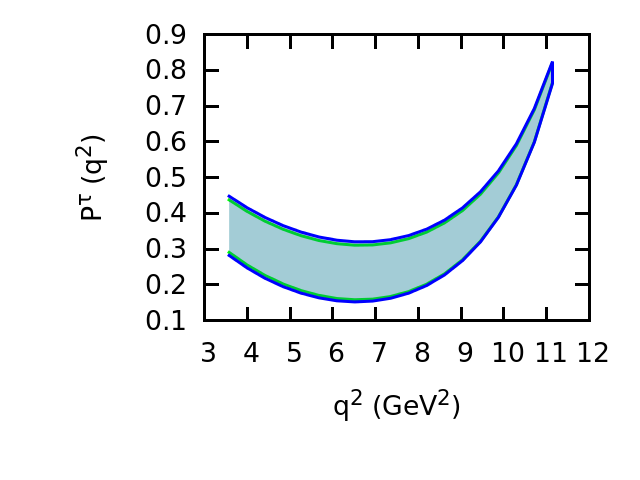}
\includegraphics[width=5.9cm,height=4.3cm]{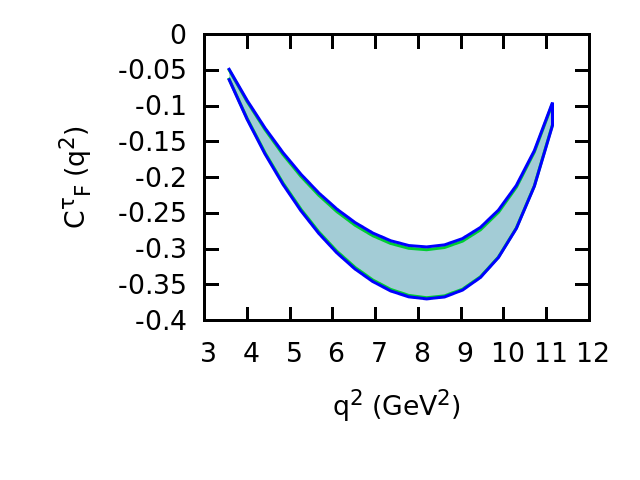}
\caption{$q^2$ dependence of various observables with the allowed ranges of $V_L$ and $V_R$ NP couplings of Fig.~\ref{vlvr} are shown with
green band. The corresponding $1\sigma$ SM range is shown with the blue band.}
\label{obs_vlvr}
\end{figure}

We show in Fig.~\ref{obs_vlvr} the $q^2$ dependence of various observables with the allowed values of $V_L$ and $V_R$ NP couplings of 
Fig.~\ref{vlvr}. The SM $1\sigma$ range is shown with blue band, whereas, the allowed range with $V_L$ and $V_R$ NP couplings is shown with
green band. It is evident from Fig.~\ref{obs_vlvr} that the differential branching ratio~${\rm DBR}(q^2)$ and ratio of branching
ratio~$R(q^2)$ deviate considerably from the SM expectation. Again, as expected, we do not observe any deviation of $A_{FB}^{\tau}(q^2)$, 
$P_{\tau}(q^2)$ and $C_{F}^{\tau}(q^2)$ from the SM expectation in this NP scenario.

\subsubsection{Scenario II: only $S_L$ and $S_R$ NP couplings}
\label{sl_sr}
\begin{figure}[htbp]
\centering
\includegraphics[width=6.0cm,height=4.3cm]{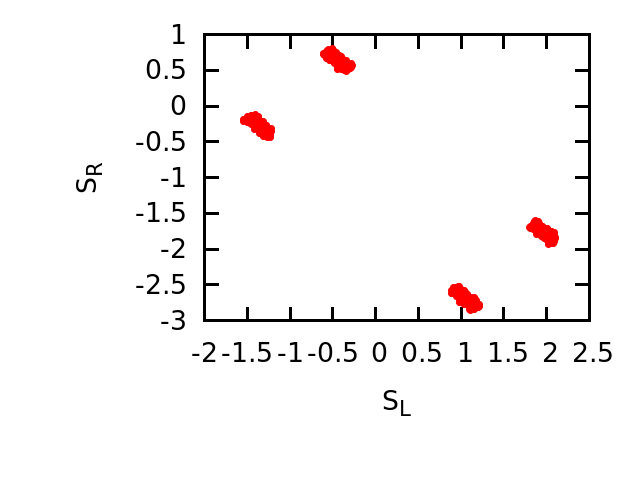}
\includegraphics[width=6.0cm,height=4.3cm]{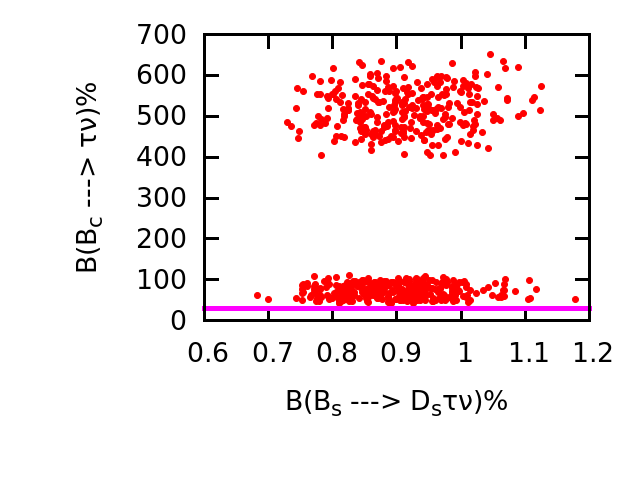}
\caption{Allowed ranges of $S_L$ and $S_R$ NP couplings are shown in the left panel once $1\sigma$ constraint coming from the measured
values of the ratio of branching ratios $R_D$ and
$R_{D^{\ast}}$ is imposed. We show in the right panel the allowed ranges in $\mathcal B(B_c \to \tau\nu)$ and $\mathcal B(B_s \to D_s\tau\nu)$
in the presence of these NP couplings. The pink constant line in the right panel corresponds to the upper bound of 
$\mathcal B(B_c \to \tau\nu) = 30\%$ obtained in the SM.}
\label{slsr}
\end{figure}

In this scenario, we consider the effect of new scalar couplings only, i.e,$(S_L,\,S_R) \ne 0$ and all the other NP couplings are zero.
In the presence of $S_L$ and $S_R$ NP couplings, the differential decay width, ratio of branching ratio, forward backward asymmetry, 
polarization fraction of the $\tau$ lepton, and the convexity parameter can be expressed as
\begin{eqnarray}
\label{slrobs}
&&\left[\frac{d\Gamma}{dq^{2}}\right]_{S_{L,R}}= \left[\frac{d\Gamma}{dq^{2}}\right]_{SM} + 8N|P_{D_s}| \left( \frac{1}{2} H_S^2\,{G}_S^2 + 
\frac{m_l}{\sqrt{q^2}}H_t\,H_S\,G_S\, \right) , \nonumber \\
&&\left[R_{D_s}\right]_{S_{L,R}}=\left[R_{D_s}\right]_{SM} + \frac{8N|P_{D_s}| \left( \frac{1}{2} H_S^2\,{G}_S^2 + 
\frac{m_l}{\sqrt{q^2}}H_t\,H_S\,G_S\, \right)}
{\mathcal B(B_s \to D_s\,e\,\nu)}\,, \nonumber \\
&&A_{FB}^l(q^2)|_{S_{L,R}}= \frac{3\,m_l^2}{2\,q^2}\left[\frac{H_0\,H_t + H_0\,\frac{\sqrt{q^2}}{m_l}\,H_S\,G_S}{H_0^2\,
\Big(1+\frac{m_l^2}{q^2}\Big) +\frac{3\,m_l^2}{2\,q^2}H_t^2 + (3/2)\,H_S^2\,G_S^2 +3\,(m_l/\sqrt{q^2})\,H_t\,H_S\,G_S}\right]\,, \nonumber \\
&&\left[P^l\right]_{S_{L,R}}= \frac{\frac{m_l^2}{2\,q^2}\,(H_0^2+3\,H_t^2) - H_0^2 + (3/2)\,H_S^2\,G_S^2 + 
3\,(m_l/\sqrt{q^2})\,H_t\,H_S\,G_S}{\frac{m_l^2}{2\,q^2} (H_0^2+3\,H_t^2) + H_0^2 + (3/2)\,H_S^2\,G_S^2 + 3\,(m_l/\sqrt{q^2})\,H_t\,H_S\,G_S}
\,, \nonumber \\
&&\left[C_F^l\right]_{S_{L,R}}= \frac{3}{2}\frac{H_0^2\left(\frac{m_l^2}{q^2}-1\right)}
{\left\{H_0^2\Big(1+\frac{m_l^2}{2q^{2}}\Big)+\frac{3\,m_l^2}{2\,q^2}\,H_t^2\right\} + (3/2)\,H_S^2\,G_S^2 + 3\,(m_l/\sqrt{q^2})\,H_t\,H_S\,
G_S}
\end{eqnarray}
We impose $1\sigma$ constraint coming from experimental values of $R_D$ and $R_{D^{\ast}}$ to determine the allowed values of $S_L$ and $S_R$
NP couplings. The resulting $(S_L,\,S_R)$ allowed ranges are shown in the left panel of Fig.~\ref{slsr}..
In the right panel, we show the corresponding ranges in $\mathcal B(B_c \to \tau\nu)$ and $\mathcal B(B_s \to D_s\tau\nu)$ obtained using the
allowed values of $(S_L,\,S_R)$ NP couplings. It should be noted that the $\mathcal B(B_c \to \tau\nu)$ obtained in this scenario is rather
large; more than $30\%$. Thus, although $(S_L,\,S_R)$ NP couplings can simultaneously explain the anomalies present in $R_D$ and 
$R_{D^{\ast}}$, it, however, fails to satisfy the $\mathcal B(B_c \to \tau\nu) \le 30\%$ constraint obtained in the SM.
\begin{table}[htbp]
\centering
\begin{tabular}{|c|c|c|c|c|}
\hline
$\mathcal B(B_s \to D_s\tau\nu)\%$ & $R_{D_s}$ & $P^{\tau}$ & $A^{\tau}_{FB}$ & $C^{\tau}_F$ \\
\hline
\hline
$[0.683, 1.178]$ & $[0.306, 0.518]$ & $[0.345, 0.609]$ & $[-0.276, 0.355]$ & $[-0.156, -0.260]$ \\
\hline
\end{tabular}
\caption{Allowed ranges of various observables in the presence of $S_L$ and $S_R$ NP couplings}
\label{tabslsr}
\end{table}
Although this particular scenario is ruled out by the $\mathcal B(B_c \to \tau\nu)$ constraint, nevertheless, we report in
Table.~\ref{tabslsr} the allowed ranges of all the observables obtained using the allowed values of $S_L$ and $S_R$
NP couplings of Fig.~\ref{slsr}. The deviation from the SM prediction observed in this scenario is quite significant. We notice that
the forward backward asymmetry parameter can assume negative values within this scenario, which is quite distinct from SM expectation.
\begin{figure}[htbp]
\centering
\includegraphics[width=5.9cm,height=4.3cm]{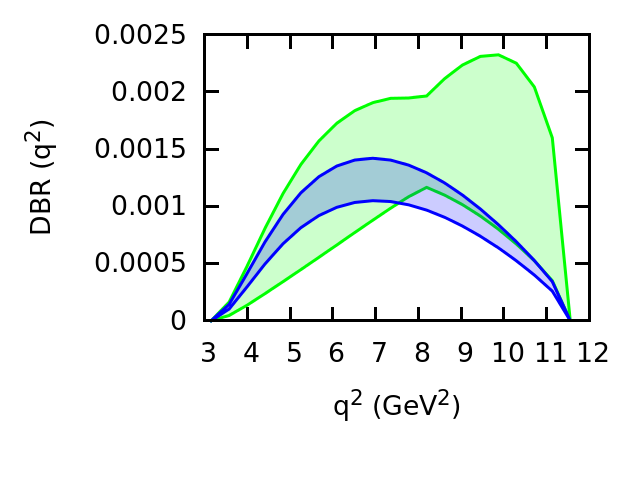}
\includegraphics[width=5.9cm,height=4.3cm]{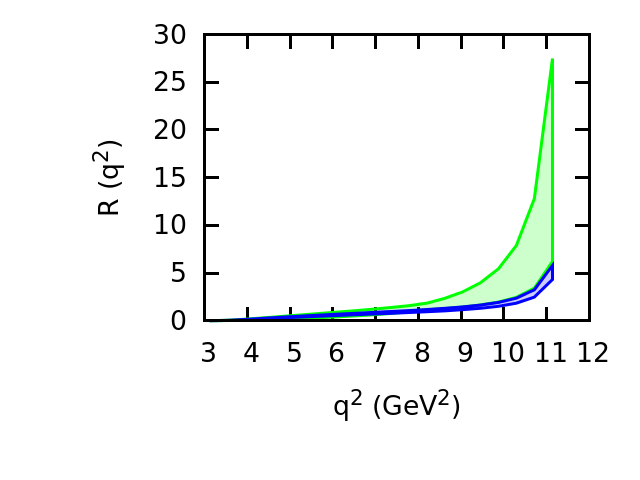}
\includegraphics[width=5.9cm,height=4.3cm]{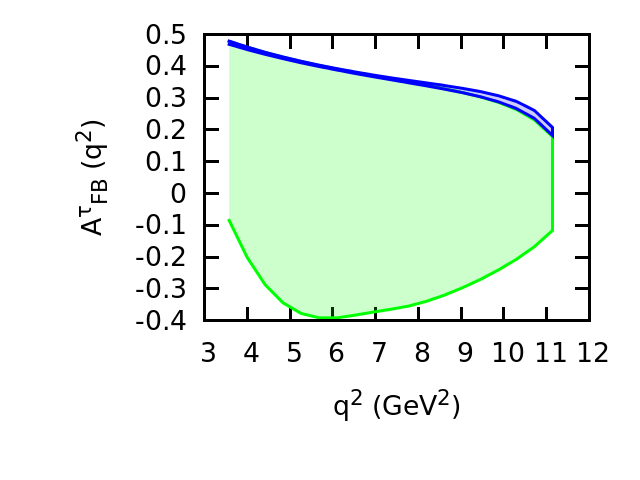}
\includegraphics[width=5.9cm,height=4.3cm]{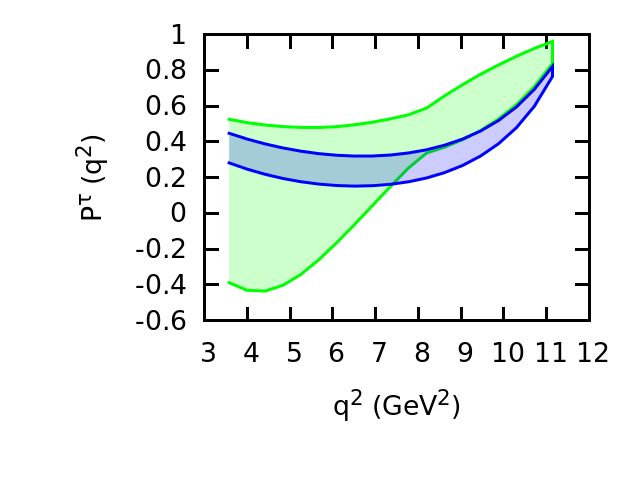}
\includegraphics[width=5.9cm,height=4.3cm]{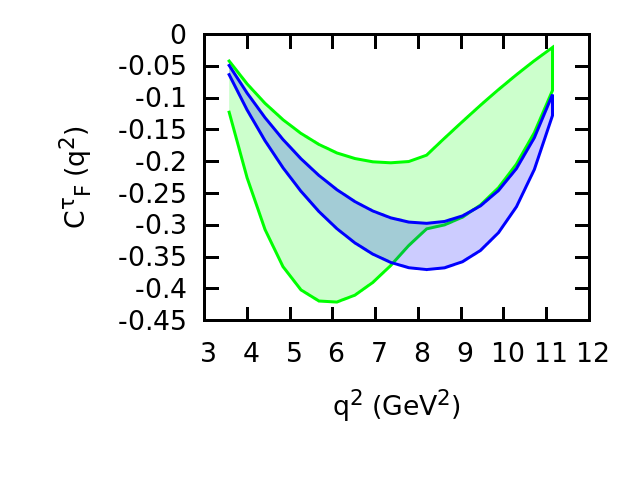}
\caption{$q^2$ dependence of various observables with the allowed ranges of $S_L$ and $S_R$ NP couplings of Fig.~\ref{slsr} are shown with
green band. The corresponding $1\sigma$ SM range is shown with the blue band.}
\label{obs_slsr}
\end{figure}

We show the effect of NP on various $q^2$ dependent observables in Fig.~\ref{obs_slsr}. We show with blue the SM $1\sigma$ band, whereas,
we show with green the allowed band once the NP is switched on. The deviation observed in this scenario is rather large and it is, indeed,
more pronounced that the deviation obtained with $(V_L,\,V_R)$ NP couplings. Unlike scenario I, there is no cancellation of NP effects in
$A_{FB}^{\tau}(q^2)$, $P^{\tau}(q^2)$, and $C_F^{\tau}(q^2)$. We notice that, although there is no zero crossing in the SM for the 
$A_{FB}^{\tau}(q^2)$ parameter, we may observe zero crossing depending on the values of $S_L$ and $S_R$ NP couplings. Similar conclusion can 
be drawn for the $\tau$ polarization fraction $P^{\tau}(q^2)$ as well. Moreover, depending on the values of the NP couplings, shape of the 
$q^2$ distribution curve of each observable can be quite different from its SM counterpart.

\subsubsection{Scenario III: only $\widetilde{V}_L$ and $\widetilde{V}_R$ NP couplings}
\label{vlt_vrt}
\begin{figure}[htbp]
\centering
\includegraphics[width=6.0cm,height=4.3cm]{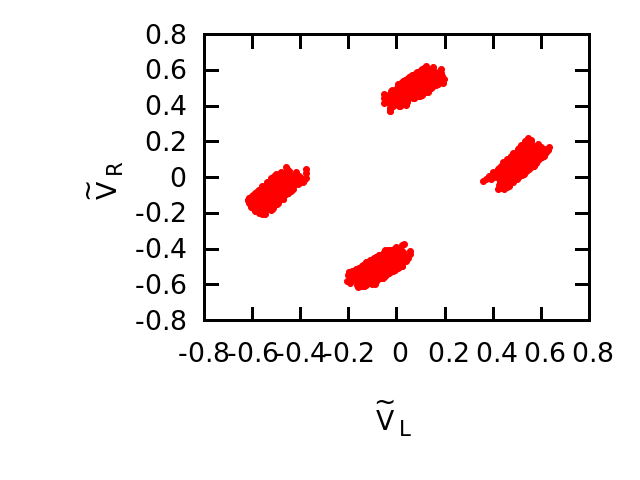}
\includegraphics[width=6.0cm,height=4.3cm]{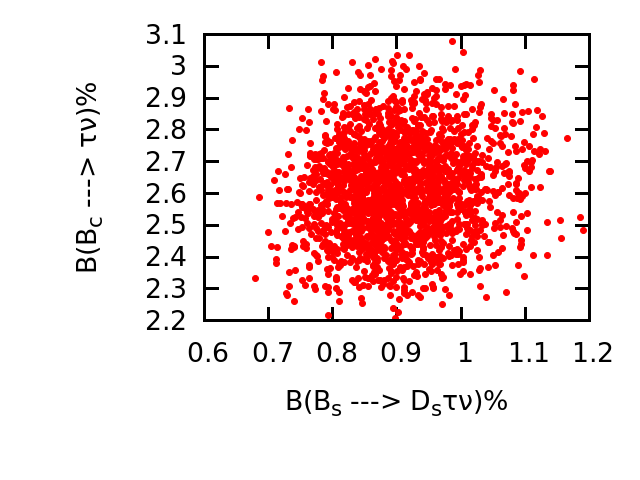}
\caption{Allowed ranges of $\widetilde{V}_L$ and $\widetilde{V}_R$ NP couplings are shown in the left panel once $1\sigma$ constraint 
coming from the measured values of the ratio of branching ratios $R_D$ and
$R_{D^{\ast}}$ is imposed. We show in the right panel the allowed ranges in $\mathcal B(B_c \to \tau\nu)$ and 
$\mathcal B(B_s \to D_s\tau\nu)$ in the presence of these NP couplings.}
\label{vltvrt}
\end{figure}

To study the effect of new vector type NP couplings associated with right handed neutrino interactions, we
consider $(\widetilde{V}_L,\, \widetilde{V}_R)$ to be nonzero while all other NP couplings to be zero.
In this scenario, the differential decay width, ratio of branching ratio, forward backward asymmetry, $\tau$ polarization fraction, and the
convexity parameter take the following simple form:
\begin{eqnarray}
\label{vltvrtobs}
&&\left[\frac{d\Gamma}{dq^2}\right]_{\widetilde{V}_{L,R}}=\left[\frac{d\Gamma}{dq^2}\right]_{SM}(1+ \widetilde{G}_V^2)\,, \qquad\qquad
\left[R_{D_s}\right]_{\widetilde{V}_{L,R}}=\left[R_{D_s}\right]_{SM}(1+\widetilde{G}_V^2)\,, \nonumber \\
&&\left[A_{FB}^l(q^2)\right]_{\widetilde{V}_{L,R}}= \left[A_{FB}^l(q^2)\right]_{SM}\,, \qquad\qquad
\left[P^l\right]_{\widetilde{V}_{L,R}}= \left[P^l\right]_{SM} \frac{(1-\widetilde{G}_V^2)}{(1+\widetilde{G}_V^2)}\,, \qquad\qquad
\left[C_F^l\right]_{\widetilde{V}_{L,R}}= \left[C_F^l\right]_{SM}
\end{eqnarray}
In the left panel of Fig.~\ref{vltvrt} we show the allowed region of $(\widetilde{V}_L$, $\widetilde{V}_R)$ NP couplings that is obtained 
once $1\sigma$ $R_D$ and $R_{D^{\ast}}$ experimental constraint is imposed. Similarly, in the right panel we show the corresponding ranges of
$\mathcal B(B_c \to \tau\nu)$ and $\mathcal B(B_s \to D_s\tau\nu)$ obtained with the $(\widetilde{V}_L$, $\widetilde{V}_R)$ NP couplings. 
Similar to Scenario I, we notice that $\mathcal B(B_c \to \tau\nu)$ obtained in this scenario lies within $(2-3)\%$ range. This is, again, 
consistent with the SM prediction.
In Table.~\ref{tabvltvrt}, we report the possible ranges of all the observables for the $B_s \to D_s\tau\nu$ decays. The deviation from the
SM prediction observed in this scenario is quite similar to the deviation observed in scenario I. However, there is one subtle difference.
Unlike scenario I, a significant deviation from the SM prediction for the $\tau$ polarization fraction is observed in this scenario. This is
evident from Eq.~\ref{vltvrtobs} that the NP effect does not get cancelled for the $\tau$ polarization fraction $P^{\tau}$. 

\begin{figure}[htbp]
\centering
\includegraphics[width=5.9cm,height=4.3cm]{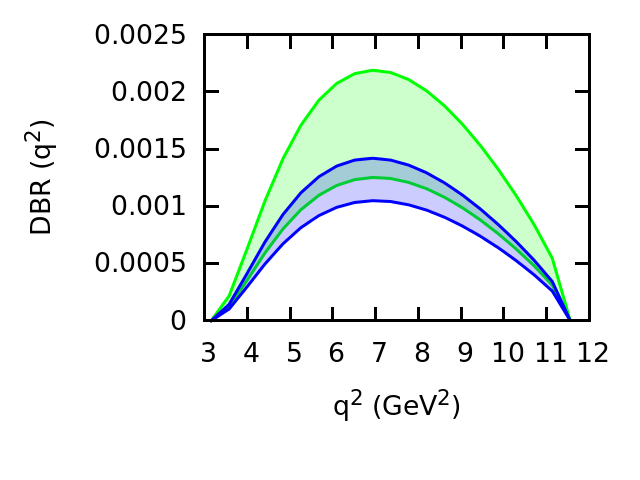}
\includegraphics[width=5.9cm,height=4.3cm]{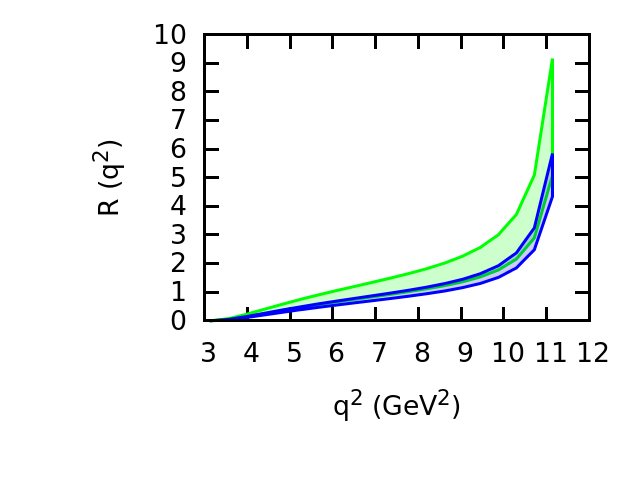}
\includegraphics[width=5.9cm,height=4.3cm]{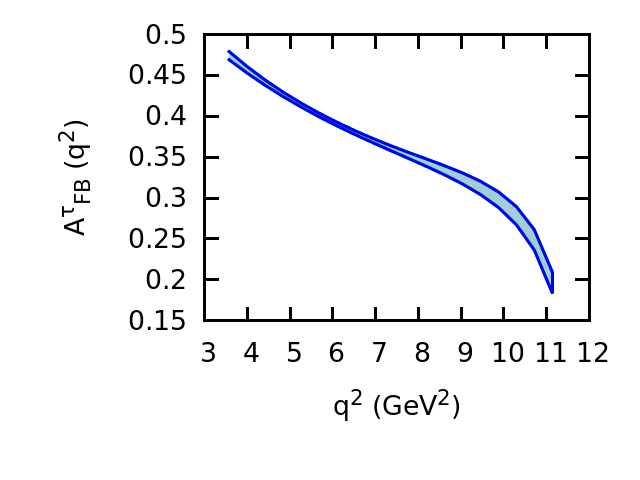}
\includegraphics[width=5.9cm,height=4.3cm]{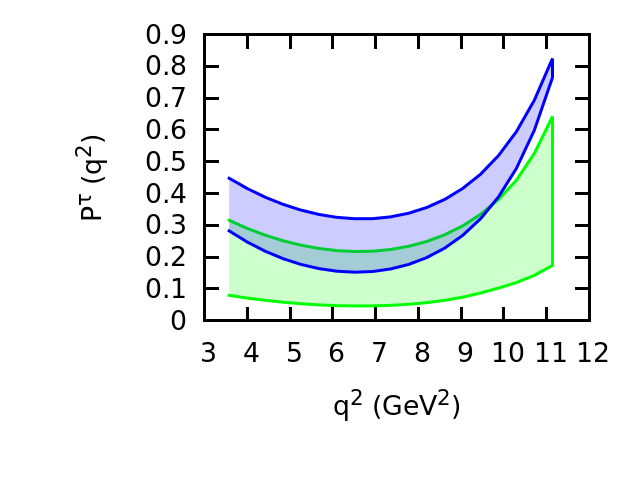}
\includegraphics[width=5.9cm,height=4.3cm]{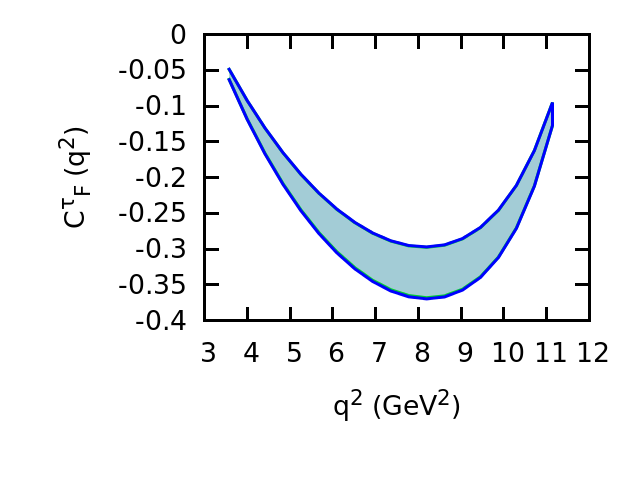}
\caption{$q^2$ dependence of various observables with the allowed ranges of $\widetilde{V}_L$ and $\widetilde{V}_R$ NP couplings of 
Fig.~\ref{vltvrt} are shown with green band. The corresponding $1\sigma$ SM range is shown with the blue band.}
\label{obs_vltvrt}
\end{figure}

In Fig.~\ref{obs_vltvrt}, we show the variation of various observables such as differential branching ratio, ratio of branching ratio, 
forward-backward asymmetry, $\tau$ polarization fraction, and convexity parameter as a function of $q^2$. The deviation from the SM
prediction observed in this scenario is quite similar to scenario I. As expected, we observe a significant deviation in $\tau$ polarization
parameter in this scenario. 
All the above mentioned analysis for the observed deviations are clearly reflected in Eq.~\ref{vltvrtobs}. 
\begin{table}[htbp]
\centering
\begin{tabular}{|c|c|c|c|c|}
\hline
$\mathcal B(B_s \to D_s\tau\nu)\%$ & $R_{D_s}$ & $P^{\tau}$ & $A^{\tau}_{FB}$ & $C^{\tau}_F$ \\
\hline
\hline
$[0.679, 1.191]$ & $[0.305, 0.525]$ & $[0.064, 0.281]$ & $[0.352, 0.364]$ & $[-0.239, -0.305]$ \\
\hline
\end{tabular}
\caption{Allowed ranges of various observables in the presence of $\widetilde{V}_L$ and $\widetilde{V}_R$ NP couplings}
\label{tabvltvrt}
\end{table}

\subsubsection{Scenario IV: only $\widetilde{S}_L$ and $\widetilde{S}_R$ NP couplings }
\begin{figure}[htbp]
\centering
\includegraphics[width=6.0cm,height=4.3cm]{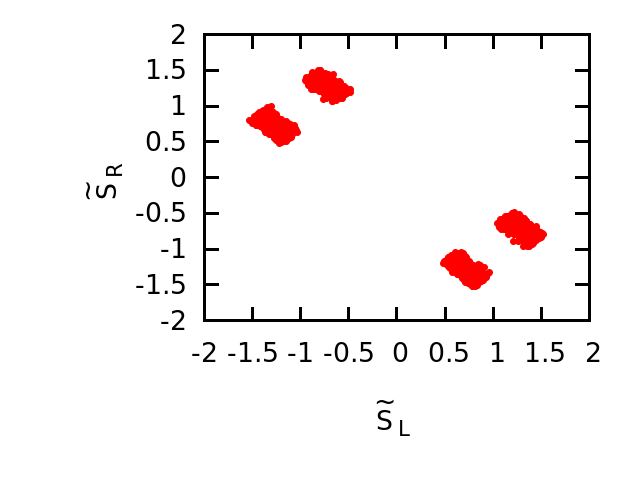}
\includegraphics[width=6.0cm,height=4.3cm]{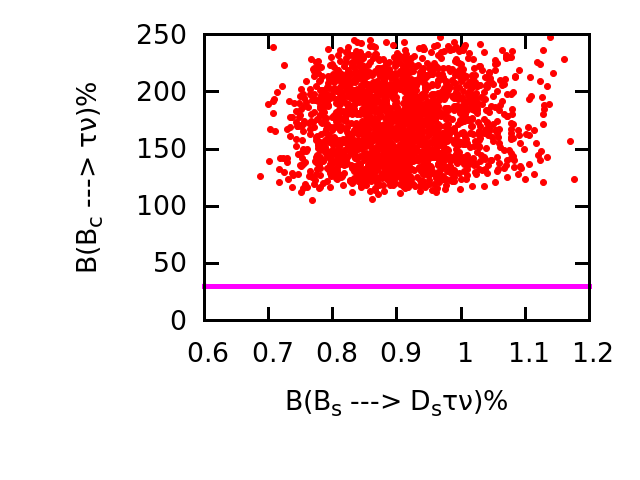}
\caption{Allowed ranges of $\widetilde{S}_L$ and $\widetilde{S}_R$ NP couplings are shown in the left panel once $1\sigma$ constraint 
coming from the measured values of the ratio of branching ratios $R_D$ and
$R_{D^{\ast}}$ is imposed. The corresponding allowed ranges in $\mathcal B(B_c \to \tau\nu)$ and $\mathcal B(B_s \to D_s\tau\nu)$
in the presence of such NP couplings are shown in the right panel. The pink constant line denotes the upper bound of 
$\mathcal B(B_c \to \tau\nu) = 30\%$ obtained in the SM.}
\label{sltsrt}
\end{figure}

In this scenario, we wish to see the effect of new scalar type NP couplings on various observables. To this end, we consider
$(\widetilde{S}_L, \widetilde{S}_R)$ to be non zero and all other NP couplings to be zero. 
In this scenario, the differential decay width, ratio of branching ratio, forward backward asymmetry, $\tau$ polarization fraction, and the
convexity parameter take the following form:
\begin{eqnarray}
\label{sltsrtobs}
&&\left[\frac{d\Gamma}{dq^2}\right]_{\widetilde{S}_{L,R}}=\left[\frac{d\Gamma}{dq^2}\right]_{SM} + 4\,N|P_{D_s}|H_S^2\,\widetilde{G}_S^2\,,
\qquad\qquad
\left[R_{D_s}\right]_{\widetilde{S}_{L,R}}=\left[R_{D_s}\right]_{SM} + \frac{(3/2)\,H_S^2\,\widetilde{G}_S^2}
{\mathcal B(B_s \to D_s\,e\,\nu)}\,, \nonumber \\
&&\left[A_{FB}^l(q^2)\right]_{\widetilde{S}_{L,R}}= \frac{3\,m_l^2}{2\,q^2}\left[\frac{H_0\,H_t}{H_0^2\,\Big(1+\frac{m_l^2}{q^2}\Big)+
\frac{3\,m_l^2}{2\,q^2}H_t^2 + (3/2)\,H_S^2\,\widetilde{G}_S^2}\right]\,, \nonumber \\
&&\left[P^l\right]_{\widetilde{S}_{L,R}}= \frac{\frac{m_l^2}{2\,q^2}(H_0^2+3\,H_t^2) - H_0^2 -(3/2)\,H_S^2\widetilde{G}_S^2}
{\frac{m_l^2}{2\,q^2}(H_0^2+3\,H_t^2) + H_0^2 +(3/2)\,H_S^2\widetilde{G}_S^2}\,, \nonumber \\
&&\left[C_F^l(q^2)\right]_{\widetilde{S}_{L,R}}= \frac{3}{2}\frac{H_0^2\left(\frac{m_l^2}{q^2}-1\right)}
{\left\{H_0^2\Big(1+\frac{m_l^2}{2\,q^2}\Big)+\frac{3\,m_l^2}{2\,q^2}\,H_t^2\right\} +(3/2)\,H_S^2\widetilde{G}_S^2}
\end{eqnarray}
In order to determine the allowed ranges of
$(\widetilde{S}_L, \widetilde{S}_R)$ NP couplings, we impose $1\sigma$ constraint coming from experimentally measured values of $R_D$ and
$R_{D^{\ast}}$. The resulting NP parameter space, shown in the left panel of Fig.~\ref{sltsrt}, can simultaneously explain the anomalies 
present in $R_D$ and $R_{D^{\ast}}$. We show in the right panel the allowed ranges in $\mathcal B(B_c \to \tau\nu)$ and
$B_s \to D_s\tau\nu$ with such NP. We notice that the $\mathcal B(B_c \to \tau\nu)$ obtained in this scenario is not compatible with the
upper bound of $\mathcal B(B_c \to \tau\nu) \le 30\%$ obtained in the SM. 
The numerical values written in the square brackets of Table-\ref{tabsltsrt} represent the allowed ranges of observables obtained with 
the allowed values of $(\widetilde{S}_L, \widetilde{S}_R)$ of Fig.~\ref{sltsrt}. Similar to scenario II, we see significant deviation of 
all the observables from the SM expectation.

\begin{figure}[htbp]
\centering
\includegraphics[width=5.9cm,height=4.3cm]{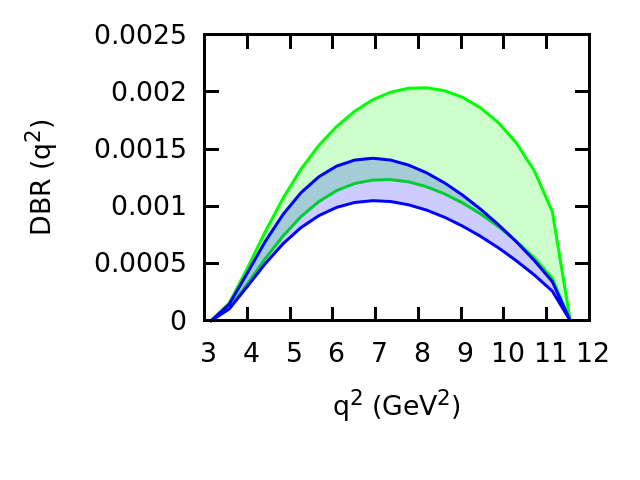}
\includegraphics[width=5.9cm,height=4.3cm]{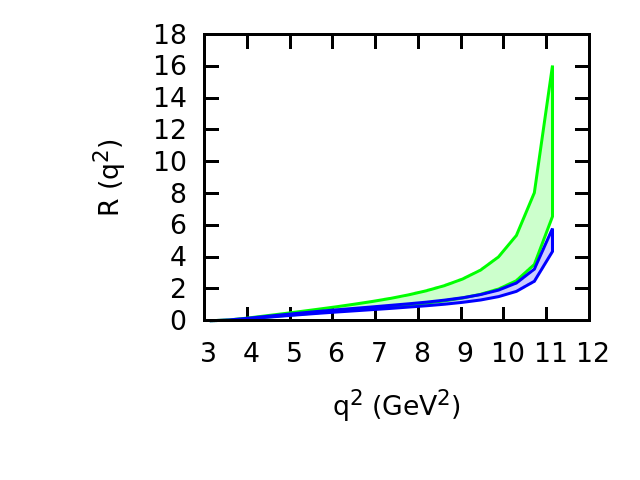}
\includegraphics[width=5.9cm,height=4.3cm]{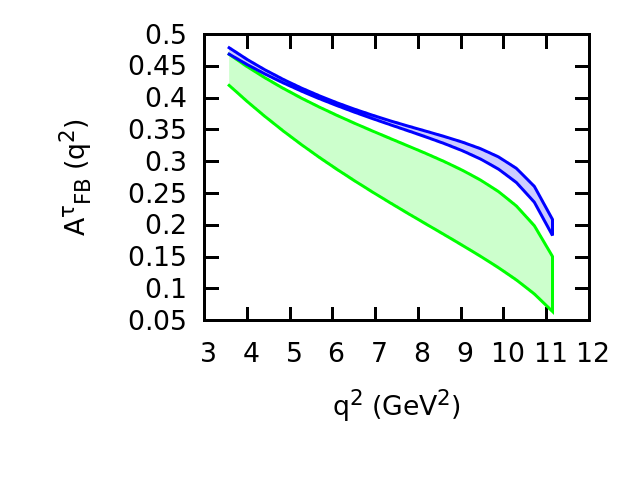}
\includegraphics[width=5.9cm,height=4.3cm]{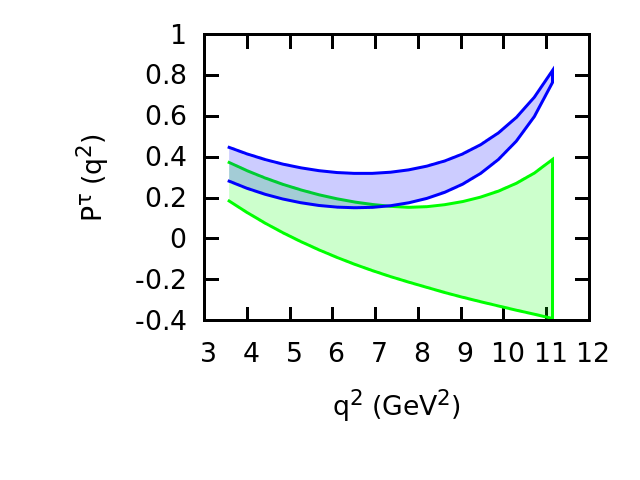}
\includegraphics[width=5.9cm,height=4.3cm]{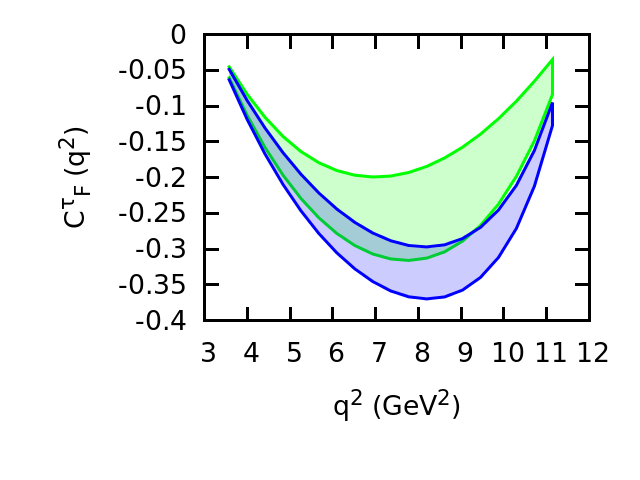}
\caption{$q^2$ dependence of various observables with the allowed ranges of $\widetilde{S}_L$ and $\widetilde{S}_R$ NP couplings of 
Fig.~\ref{sltsrt} are shown with green band. The corresponding $1\sigma$ SM range is shown with the blue band.}
\label{obs_sltsrt}
\end{figure}

We show in Fig.~\ref{obs_sltsrt} the $q^2$ distribution of various observables for the $B_s \to D_s\tau\nu$ decays. The blue band
corresponds to the $1\sigma$ SM range, whereas, the green band corresponds to the range of the observables once the 
$(\widetilde{S}_L, \widetilde{S}_R)$ NP couplings are switched on. The deviation observed in this scenario is rather large. We notice that
although, in the SM, there is no zero crossing in the $\tau$ polarization parameter, there may or may not be a zero crossing depending on
the values of the NP couplings. For the differential branching ratio, the peak of the $q^2$ distribution may shift towards high $q^2$
region.
\begin{table}[htbp]
\centering
\begin{tabular}{|c|c|c|c|c|}
\hline
$\mathcal B(B_s \to D_s\tau\nu)\%$ & $R_{D_s}$ & $P^{\tau}$ & $A^{\tau}_{FB}$ & $C^{\tau}_F$ \\
\hline
\hline
$[0.687, 1.176]$ & $[0.305, 0.532]$ & $[-0.201, 0.205]$ & $[0.219, 0.331]$ & $[-0.155, -0.261]$ \\
\hline
\end{tabular}
\caption{Allowed ranges of various observables in the presence of $\widetilde{S}_L$ and $\widetilde{S}_R$ NP couplings}
\label{tabsltsrt}
\end{table}

\section{Conclusion}
\label{con}
In view of the long standing anomalies in $R_D$ and $R_{D^{\ast}}$, we study the corresponding $B_s \to D_s\tau\nu$ semileptonic decays
in a model independent framework. We use the helicity formalism to study the $B_s \to D_s\tau\nu$ semileptonic decays within the context 
of an effective Lagrangian in the presence of NP and 
explore four different NP scenarios based on contributions coming from two different NP operators. We give prediction on various observables
such as branching ratio, ratio of branching ratio, forward backward asymmetry, longitudinal polarization fraction of the charged lepton, and 
the convexity parameter for this decay mode within SM and within four different NP scenarios. 

We first report the central values and the $1\sigma$ ranges of each observable within the SM for both the $e$ and the $\tau$ modes. We notice
that all the observables change considerably while going from the $e$ mode to the $\tau$ mode. The value of $R_{D_s}$ is quite similar to
the value reported in Ref.~\cite{Monahan:2017uby}. We also give the first prediction of the longitudinal polarization fraction of the 
charged lepton, lepton side forward backward asymmetry, and the convexity parameter for the $B_s \to D_s\,l\,\nu$ decays. 

For our NP analysis, we assume that NP effects are coming from  vector and scalar type NP couplings only. We notice that NP scenarios
with $(V_L,\,V_R)$ and $(\widetilde{V}_L,\,\widetilde{V}_R)$ NP couplings are compatible with the $\mathcal B(B_c \to \tau\nu)$ constraint.
However, NP scenarios with $(S_L,\,S_R)$ and $(\widetilde{S}_L,\,\widetilde{S}_R)$ NP couplings are ruled out due to the constraint coming 
from the lifetime of $B_c$ meson.  

Study of $B_s \to D_s\tau\nu$ decays both theoretically and experimentally is crucial because it may provide new insights into the $R_D$ 
and $R_{D^{\ast}}$ anomaly as this decay mode is mediated via the same $b \to c$ charged current interaction.
Moreover, a precise determination of the branching fractions of this decay mode will allow an accurate determination of the CKM matrix 
element $|V_{cb}|$.

\end{document}